\documentclass[12pt]{article}
\usepackage{ametsoc}
\linenumbers
\newcommand{\myabstract}{This paper presents a simple analytical framework for the dynamic response of cirrus to a local radiative flux convergence, expressible in terms of three independent modes of cloud evolution. Horizontally narrow and tenuous clouds within a stable environment adjust to radiative heating by ascending gradually across isentropes while spreading sufficiently fast so as to keep isentropic surfaces nearly flat. More optically dense clouds experience very concentrated heating, and if they are also very broad, they develop a convecting mixed layer. Along isentropic spreading still occurs, but in the form of turbulent density currents rather than laminar flows. A third adjustment mode relates to evaporation, which erodes cloudy air as it lofts. The dominant mode is determined from two dimensionless numbers, whose predictive power is shown in comparisons with high resolution numerical cloud simulations. The power and simplicity of the approach hints that fast, sub-grid scale radiative-dynamic atmospheric interactions might be efficiently parameterized within slower, coarse-grid climate models.}
\begin{document}
%
%
\title{\textbf{\large{A Simple Framework For the Dynamic Response of Cirrus Clouds to Local Diabatic Radiative Heating}}}
%
%
\author{\textsc{Clinton T. Schmidt and Timothy J. Garrett}
				\thanks{\textit{Corresponding author address:} 
				Tim Garrett, Department of Atmospheric Sciences, University of Utah 
				135 S 1460 E RM 819 (WBB), Salt Lake City, UT 84112-0110. 
				\newline{E-mail: tim.garrett@utah.edu}}\\
\textit{\footnotesize{University of Utah, Department of Atmospheric Sciences, Salt Lake City, Utah}}
}
%
\ifthenelse{\boolean{dc}}
{
\twocolumn[
\begin{@twocolumnfalse}
\amstitle

\begin{center}
\begin{minipage}{13.0cm}
\begin{abstract}
	\myabstract
	\newline
	\begin{center}
		\rule{38mm}{0.2mm}
	\end{center}
\end{abstract}
\end{minipage}
\end{center}
\end{@twocolumnfalse}
]
}
{
\amstitle
\begin{abstract}
\myabstract
\end{abstract}
\newpage
}
\section{Introduction}

Cloud-climate feedbacks remain a primary source of uncertainty in
climate forecasts \citep{Dufresne&Bony08}, mainly because clouds
both drive and respond to the general circulation, the hydrological
cycle, and the atmospheric radiation budget. Unlike fields of water
vapor, clouds evolve quickly, so their radiative forcing and dynamic
evolution are highly coupled on time and spatial scales that cannot
be easily resolved within Global Climate Models (GCMs). For faithful
reproduction of large-scale climate features, resolving radiatively
driven motions on sub-grid scales may be at least as important as
accurately representing mean grid-scale fluxes \citep{cole2005}. 

Radiative flux convergence and divergence within cloudy air is normally
thought to produce vertical lifting and mixing motions \citep{Danielsen82,Ackerman88,Lilly88,JensenEtAl96,DobbieAndJonas2001}.
What is often overlooked is that clouds with a finite width also adjust
to radiative heating by spreading horizontally, especially if the
heating is concentrated in a thin layer at the cloud top or bottom
\citep{GarrettEtAl2005,Garrett06}. Such radiatively driven mesoscale
circulations have been identified within thin tropopause cirrus, and
they are thought to play a role in determining the heating rate of
the upper troposphere \citep{Durran09} and in stratospheric dehydration
mechanisms \citep{Dinh&Durran10}. \citet{JensenTTLCirrusDynamics}
suggest that radiative cooling can help to initiate thin tropopause
cirrus formation, while subsequent radiative heating in an environment
of weak stability can induce the small-scale convection currents that
are required to maintain the cloud against gravitational sedimentation
and vertical wind shear.

Where these recent studies directly simulated the highly interactive
and complex nature of cloud processes, an alternative and perhaps
more general approach is to start with simple, analytical and highly
idealized models that emphasize specific aspects of the relevant physics.
Here, we look at the respose of cirrus clouds to local thermal radiative
flux divergence within cloud condensate. The discussion that follows
largely neglects precipitation, synoptic scale motions, and shear
dynamics to facilitate description of a simple theoretical framework
within a parameter space of two dimensionless numbers. A similar approach
has been employed previously to constrain small-scale interactions
between diabatic heating and atmospheric dynamics \citep{Raymond&Rotunno89},
including situations where radiation is absorbed by horizontally infinite
clouds \citep{DobbieAndJonas2001}. Here, we extend consideration
to radiatively absorptive layers that have finite lateral dimensions,
an ingredient that turns out to be critical for predicting the evolution
of cloud size and cross-isentropic motions. The broad intent of this
study is to provide insight into how clouds respond to rapid, small-scale
radiative heating in a way that  might be better parameterized within large scale,
coarse-grid models such as GCMs.

\section{Non-equilibrium radiative-dynamic interactions in cirrus}

\begin{figure}[t]
\begin{centering}
\noindent\includegraphics[width=0.8\paperwidth,angle=0]{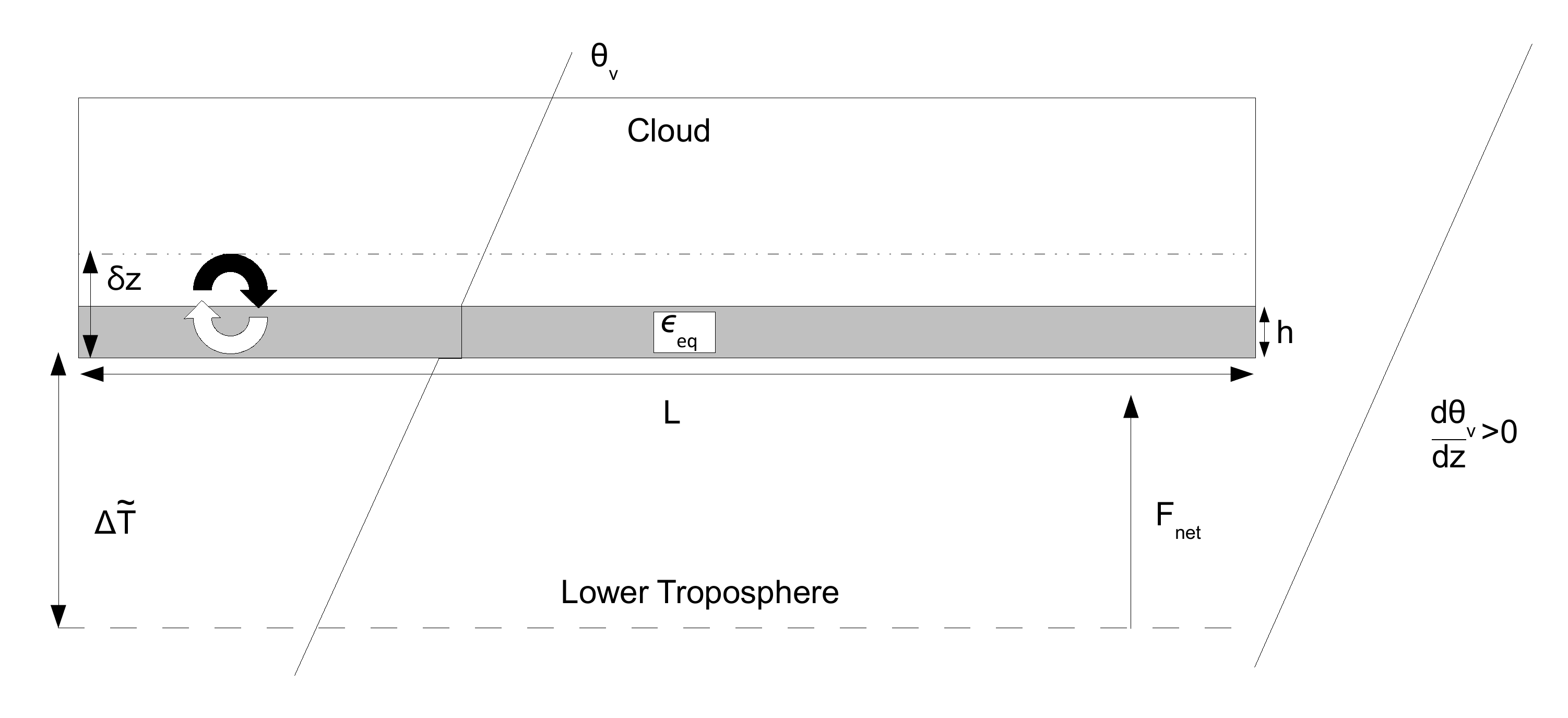}\\
\par\end{centering}

\caption{Radiative energy is transferred from the lower troposphere to the
base of a gravitationally stratified cloud with initial width $L$,
due to a radiative temperature difference $\Delta\widetilde{T}$ and
deposited into a layer of characteristic depth $h$ at the base of
the cloud. The layer is initially at equilibrium buoyant potential
$\epsilon_{eq}$ with respect surrounding clear air at the same level,
and it is perturbed from equilibrium through the deposition of radiative
energy into a well-mixed layer of depth $\delta z$.\label{fig:Diagram-of-initial}}
\end{figure}

The starting point is to consider a microphysically uniform, optically
opaque cloud that is initially at rest with respect to its surrounding, characterized by a 
stably stratified atmosphere with a virtual potential temperature
$\theta_{v}$ that increases monotonically with height (Figure \ref{fig:Diagram-of-initial}).
The arguments described below apply equally to cloud base and cloud
top, differing only in sign of forcing. However, for the sake of simplicity
the focus here is on cloud base. 

At cloud base, cloudy air has a lower brightness temperature than
the brightness temperature of the ground and lower tropospheric air
that is below it. This radiative temperature difference drives a net
flow of radiative energy into the colder cloud base, effectively due
to a gradient in photon pressure, that can be approximated as \begin{equation}
\Delta F_{net}\simeq4\sigma\widetilde{T}_{c}^{3}\Delta\widetilde{T}\end{equation}
where $\sigma$ is the Stephan-Boltzmann constant, $\widetilde{T}_{c}$
is the cloud temperature, and $\Delta\widetilde{T}$ is the effective
brightness temperature difference between the lower tropospheric air
and cloud base.

Provided the cloud is sufficiently opaque to act as a blackbody, radiative
energy is deposited within a layer of characteristic depth $h$ at
the base of the cloud that is smaller than the depth of the cloud
itself. The magnitude of $h$ can be obtained by considering that
the thermal emissivity is given by \begin{equation}
\varepsilon\simeq1-\exp(-\tau_{abs}/\bar{\mu})\end{equation}
where $\tau_{abs}$ is the absorption optical depth and $\bar{\mu}$
is the quadrature cosine for estimating the integrated contribution
of isotropic radiation to vertical fluxes. Usually $\bar{\mu}\sim0.6$
\citep{Herman1980}. The absorption optical depth is determined by
the cloud ice mixing ratio $q_{i}$, as well as the ice crystal effective
radius $r_{e}$, through $\tau_{abs}=k(r_{e})q_{i}\rho\Delta z$ where
$k$ is the mass specific absorption cross-section density, $\rho$
is the density of air, and $\Delta z$ is the vertical path length
through which the radiation is absorbed. The depth $h$ is the e-folding
path length for the attenuation such that $\tau_{abs}/\bar{\mu}=1$:\begin{equation}
h=\frac{\bar{\mu}}{k(r_{e})q_{i}\rho}\label{eq:penetration depth}\end{equation}
Assuming an effective radius of 20 $\mu$m, the value for $k(r_{e})$
in cirrus is approximately 0.045 m$^{2}$g$^{-1}$\citep{Knollenberg93}.
Taking, for example, $q_{i}$ values of 1 g kg$^{-1}$ that have been
observed in medium sized cirrus anvils in Florida \citep{GarrettEtAl2005},
the depth $h$ would be about 30 m. As a contrasting example, a cloud
with $q_{i}$ values of 0.01 g kg$^{-1}$, similar to those observed
in thin cirrus \citep{thinCirrus}, would have a radiative penetration
depth $h$ of about 3000 m. Thus, the deposition of radiative enthalpy
in this layer increases its temperature at rate \begin{equation}
\mathcal{H}=\frac{d\theta_{v}}{dt}=\frac{-1}{\rho c_{p}}\frac{dF}{dz}\simeq\frac{\Delta F_{net}}{\rho c_{p}h}=4\sigma\widetilde{T}_{c}^{3}\frac{k(r_{e})q_{i}}{\bar{\mu}c_{p}}\Delta\widetilde{T}\label{eq:heating rate}\end{equation}
where $c_{p}$ is the specific heat of the air. To first order, heating
rates are proportional to the radiative temperature contrast and the
cloud ice mixing ratio.

\subsection{Dynamic Adjustment to Diabatic Heating}

The \emph{total} flow of upwelling radiative energy into a cloud is
proportional to $\Delta F_{net}$ and the normal cloud horizontal
cross-section, which is of order $L^{2}$ where $L$ is the cloud
horizontal width. Defining the initial, neutrally buoyant, ground-state
for the gravitational potential energy density of the cloudy air within
the volume $hL^{2}$ as $\epsilon_{eq}=E_{eq}/\left(hL^{2}\right)$
(Figure \ref{fig:Diagram-of-initial}), then an accumulated flow of
energy into the volume increases the gravitational potential energy
density to $\epsilon_{eq}+\Delta\epsilon$ at rate $d\Delta\epsilon/dt=R\Delta F_{net}/(c_{p}h)$,
where $R$ is the gas constant for air. The remaining fraction ($c_{p}-R)/c_{p}=c_{v}/c_{p}$
of the radiative enthalpy deposited in the cloud goes towards increasing
the rotational and translational energy density of the cloudy air
within the layer. Conceptually, it is useful to consider the gravitational
increase as an increase in the pressure gradient that is available
to drive fluid dynamic motions: pressure gradients have units of energy
density.

\begin{figure}[t]
\begin{centering}
\includegraphics[width=0.8\paperwidth,angle=0]{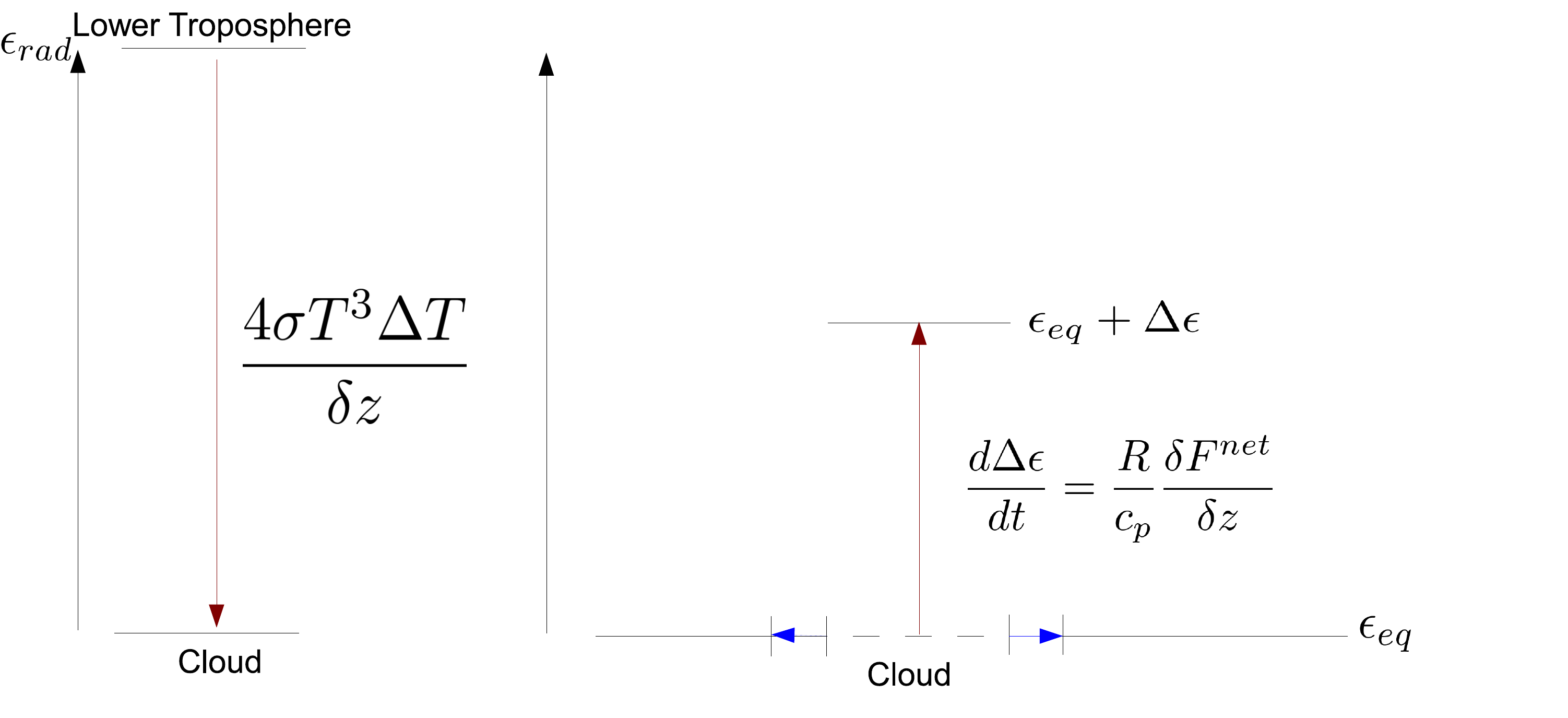}
\par\end{centering}

\caption{\label{fig:schematic}A Schematic diagram of the thermodynamic evolution
of a cirrus cloud in response to radiative diabatic heating. Potential
energy flows from the warmer lower troposphere into the cooler cloud
base (red arrow, left). The potential difference between the cloud
and the ground is $\Delta\epsilon\simeq\frac{4}{c}\sigma T^{3}\Delta T$
where $\Delta T$ is the effective brightness temperature difference
between the cloud and the lower troposphere. This flow of radiative
potential energy perturbs the cloud from gravitational equilibrium
at a rate $\frac{d\Delta\epsilon}{dt}$ (red arrow, right). The cloud
acts to restore gravitational equilibrium at rate $\alpha\Delta\epsilon$,
resulting in horizontal spreading of the cloud (blue arrows).}
\end{figure}

The increase in the potential energy density within the volume $hL^{2}$
allows work to be done against the overlying gravitational static
stability to create a mixed-layer with, on average, near constant
$\theta_{v}$. Thus any newly absorbed thermal energy becomes redistributed
through a mixed layer depth $\delta z$ that is larger than the radiatively
absorbing layer of depth $h$ (Figure \ref{fig:Diagram-of-initial}).
This is important, because it has the effect of diluting the density
of newly added radiative energy through a factor of $\delta z/h$
such that:\begin{equation}
\frac{d\Delta\epsilon}{dt}=\frac{R\Delta F_{net}}{c_{p}\delta z}\end{equation}
As required by the second law of thermodynamics, equilibrium is restored
through relaxation of the buoyant potential energy density perturbation
$\Delta\epsilon$ to zero, leading to kinematic flows (Figure \ref{fig:schematic}). 

There are two basic modes for relaxation of the buoyant energy density.
The available gravitational potential energy density $\Delta\epsilon$
can be expressed as the density $\rho$ of the air at a given buoyant
potential energy density, multiplied by the buoyant potential per
unit mass of air that is available to drive flows \begin{equation}
\Delta E\sim N^{2}\delta z^{2}\label{eq:N^2dz^2}\end{equation}
Here, $N$ is the buoyancy frequency, which is related to the local
stratification through \begin{equation}
N^{2}=\frac{g}{\theta_{v}}\frac{d\theta_{v}}{dz}\label{eq:Nsquared}\end{equation}
It follows that \begin{equation}
\Delta\epsilon=\rho N^{2}\delta z^{2}\end{equation}

Dynamic relaxation of the radiatively induced perturbation $\Delta\epsilon$
can proceed in either of two ways. At constant density, the heated
volume $L^{2}\delta z$ can be raised to higher gravitational potential.
Alternatively, the air expands outwards along a constant potential
surface.\begin{equation}
\frac{d\ln\Delta\epsilon}{dt}=\frac{\partial\ln\Delta E}{\partial t}\mid_{\rho}+\frac{\partial\ln\rho}{\partial t}\mid_{\Delta E}\label{eq:time-evolv-base}\end{equation}
Given Eq. \ref{eq:N^2dz^2} and that\begin{equation}
\rho=\frac{m}{V}\sim\frac{m}{L^{2}\delta z}\end{equation}
where $m/\delta z$ is fixed (i.e., no entrainment of mass across
the mixed-layer boundary), Eq. \ref{eq:time-evolv-base} can be rewritten
as \begin{align}
\frac{d\ln\Delta\epsilon}{dt} & =2\frac{\partial\ln\delta z}{\partial t}\mid_{L}-2\frac{\partial\ln L}{\partial t}\mid_{\delta z}\label{eq:time-evolv-mix/spread}\\
= & \alpha_{\delta z}-\alpha_{L}\end{align}
where $\alpha_{\delta z}$ and $\alpha_{L}$ represent instantaneous
rates of adjustment.

Eq. \ref{eq:time-evolv-mix/spread} has several implications. The
buoyant potential energy density $\Delta\epsilon$ within the mixed-layer
volume $L^{2}\delta z$ can grow due to the continuing radiative flux
deposition within the volume (the positive first term in Eq. \ref{eq:time-evolv-mix/spread}).
Or, it can decay through horizontal expansion (the negative second
term in Eq. \ref{eq:time-evolv-mix/spread}). In the first case, if
the width of the cloud $L$ is held constant, the mixed-layer deepens
into stratified cloudy air above it at rate \begin{equation}
\frac{\partial(\delta z)}{\partial t}\mid_{L}=\frac{d\theta_{v}/dt}{d\theta_{v}/dz}=\frac{\mathcal{H}gh}{\theta_{v}N^{2}\delta z}\label{eq:mix-layer-depth}\end{equation}
where the factor of $h/\delta z$ arises from the dilution of potential
energy through a depth larger than the absorptive layer where initially,
$\delta z=h$. Mixed-layer growth rates slow with time. The solution
to Eq. \ref{eq:mix-layer-depth} as a function of time $\Delta t$
is \begin{equation}
\delta z=(\frac{\mathcal{H}gh}{\theta_{v}N^{2}}\Delta t)^{1/2}\label{eq:t-one-half}\end{equation}

Alternatively, if the depth of the mixed-layer $\delta z$ is fixed,
then the potential energy density relaxes towards equilibrium by smoothing
out horizontal pressure gradients between the cloudy mixed-layer and
clear sky beside it. It does this through expansion of the volume
$L^{2}\delta z$ along constant potential surfaces (or isentropes)
into the lower potential energy density environment that surrounds
the cloud. This density current outflow occurs at speed \begin{equation}
u_{mix}=\frac{\partial L}{\partial t}\mid_{\delta z}\sim N\delta z\label{eq:u_mix}\end{equation}
which results from the conversion of the gravitational potential energy
of order $N^{2}\delta z^{2}$ into kinetic energy of order $u_{mix}^{2}$.

To assess the relative importance of cross-isentropic adjustment to
along-isentropic spreading, a dimensionless number $S$ can be defined
as the ratio of the two rates $\alpha_{\delta z}$ and $\alpha_{L}$
in Eq. \ref{eq:time-evolv-mix/spread}. From Eq. \ref{eq:mix-layer-depth},
radiative heating increases the mixed-layer gravitational potential
energy density at rate \begin{equation}
\alpha_{\delta z}\sim2\frac{\partial\ln\delta z}{\partial t}\mid_{L}=\frac{2\mathcal{H}gh}{\theta_{v}N^{2}\delta z^{2}}\label{eq:alpha-delta-z}\end{equation}
From Eq. \ref{eq:u_mix}, the rate of loss of potential energy density
due to expansion of the mixed-layer laterally into the clear-sky surroundings
is \begin{equation}
\alpha_{L}\sim2\frac{\partial\ln L}{\partial t}\mid_{\delta z}=2\frac{N\delta z}{L}\label{eq:alpha-L}\end{equation}

For a cloud that is initially at rest, in which case a mixed layer
has not yet developed, then radiation deposition remains concentrated
within the layer $\delta z\sim h$ and from Eqs. \ref{eq:alpha-delta-z}
and \ref{eq:alpha-L}, the ratio of these two rates can be defined
by a dimensionless {}``Spreading Number'' \begin{equation}
S=\frac{\alpha_{\delta z}}{\alpha_{L}}=\frac{\partial\ln\delta z/\partial t\mid_{L}}{\partial\ln L/\partial t\mid_{\delta z}}=\frac{\mathcal{H}gL}{\theta_{v}N^{3}h^{2}}\label{eq:Spreading-Number}\end{equation}
If $S>1$, then the potential energy density within the layer $L^{2}\delta z$
increases due to radiative flux deposition at a rate $\alpha_{\delta z}$
that is faster than the rate $\alpha_{L}$ at which gravitational
relaxation can reduce the disequilibrium in potential energy density
through horizontal flows into surrounding clear air. Isentropic surfaces
at cloud base cannot stay flat, but rather are deformed downward by
the radiative heating. This deformation creates a deepening turbulent
mixed layer that gradually grows into the overlying static stability
of the atmosphere as the square root of time (Eq. \ref{eq:t-one-half}).
Meanwhile, the mixed-layer spreads outward along isentropic surfaces
at rate $u_{mix}$ (Eq. \ref{eq:u_mix}). 

By contrast, when $S<1$, adjustment through isentropic spreading
is sufficiently rapid that isentropic surfaces stay approximately
flat. Cloud motions stay laminar rather than becoming turbulent. The
mixed-layer horizontal expansion given by $\alpha_{L}$ (Eq. \ref{eq:alpha-L})
decreases the potential energy density faster than the rate $\alpha_{\delta z}$
(Eq. \ref{eq:alpha-delta-z}) at which potential energy density is
deposited at cloud base through radiative flux convergence. The potential
energy density at cloud base does not increase and does not overcome
the overlying static stability. Rather the cloud simply lofts across
isentropic surfaces at speed \begin{equation}
w_{strat}=\frac{\mathcal{H}}{d\theta_{v}/dz}=\frac{\mathcal{H}g}{\theta_{v}N^{2}}\end{equation}
Dimensional continuity arguments require that the cloud spreads laterally
along isentropes at speed \begin{equation}
u_{strat}\sim w_{strat}\frac{L}{h}=\frac{\mathcal{H}gL}{\theta_{v}N^{2}h}\end{equation}

\subsection{Evaporative Adjustment}

The above describes two modes for how radiative flux deposition can
create pressure, or potential energy density gradients that drive
cloud-scale motions. A third possibility for adjustment is that local
radiative heating may result in microphysical changes where temperature
is maintained, but condensate evaporates or condenses. 

Assuming that all absorbed radiative energy goes towards evaporation
at cloud base, and that there is no lag associated with the diffusion
of vapor away from ice crystals, then ice evaporates at rate $\rho L_{s}dq_{i}/dt=\Delta F_{net}/h$,
where $L_{s}$ is the latent heat of sublimation. Substituting Eq.
\ref{eq:penetration depth} for $h$, radiative heating evaporates
cloud base at rate \begin{equation}
\alpha_{evap}=\left|\frac{d\ln q_{i}}{dt}\mid_{T}\right|=\frac{k(r_{e})\left|\Delta F_{net}\right|}{\bar{\mu}L_{s}}\label{eq:alpha_evap}\end{equation}
Note that if there were net radiative flux divergence, as might be
expected at the top of a thermally opaque cloud, then net cooling
would lead to condensation. 

The ratio of $\alpha_{evap}$ (Eq. \ref{eq:alpha_evap}) to $\alpha_{\delta z}$
(Eq. \ref{eq:alpha-delta-z}) implies a dimensionless {}``Evaporation
Number'' comparing the evaporation rate to the rate of laminar adjustment
through cross isentropic ascent. In the initial stages of development,
where $\delta z=h$,\begin{equation}
E=\frac{\alpha_{evap}}{\alpha_{\delta z}}=\frac{\theta_{v}N^{2}h}{2g\mathcal{H}}\frac{k(r_{e})\left|\Delta F_{net}\right|}{\bar{\mu}L_{s}}\end{equation}
or, substituting Eq. \ref{eq:Nsquared} and Eq. \ref{eq:heating rate}
for the heating rate $\mathcal{H}$

\begin{equation}
E=\frac{c_{p}h}{L_{s}q_{i}}\frac{d\theta_{v}}{dz}\label{eq:Evaporation Number}\end{equation}
The susceptibility to evaporation depends only on the cloud microphysics
and the local static stability, and not, in fact, on the magnitude
of the heating. Provided $E>1$, cloud base evaporates rather than
lofts. However, for values of $E<1$, cloud ascends faster than it
evaporates and condensate is maintained. It is important to note here
that the {}``Evaporation Number'' $E$ should only be considered
if the {}``Spreading Number'' $S$ has values smaller than unity.
If $S>1$, the relevance of evaporation is less clear because a convective
mixed-layer develops, in which case one would expect instead continual
reformation and evaporation of cloud condensate as part of localized
circulations within the mixed-layer. The more relevant comparison
might be to rates of turbulent entrainment and mixing.

\section{Numerical Model}

To test the suitability of the dimensionless {}``Spreading'' and
{}``Evaporation'' numbers $S$ and $E$ for determining the cloud
evolutionary response to local diabatic heating, \textcolor{black}{we
made comparisons to cloud simulations from the University of Utah
Large Eddy Simulation Model (UU LESM) \citep{Zulauf}. An LES model
is used because the resolved scales are sufficiently small to represent
turbulent motions, convection, entrainment and mixing, and laminar
flows. }

The UU LESM is based on a set of fully prognostic
3D non-hydrostatic primitive equations that use the quasi-compressible
approximation \citep{Zulauf}. The model domain was placed at the
equator, $\phi=0^{\circ}$, to eliminate any Coriolis effects. Even
in the largest domain simulations, the maximum departure from the
equator (50 km) is sufficiently small as to justify not including
the Coriolis effect in the model calculations.

The horizontal extent of the domain was chosen to
contain the initialized cloud as well as to allow sufficient space
for spreading of the cloud during the model run. The UU LESM model
employs periodic boundary conditions such that fluxes through one
side of the domain (moisture, cloud ice, turbulent fluxes, etc.) enter
back into the model domain from the opposite side. Here, the horizontal
domain size is case dependent but chosen to be sufficiently large
as to minimize {}``wrap around'' effects. Horizontal grid size was
chosen to be 30 m to match the minimum value for vertical penetration
depth of radiation into the cloud, but it increased to 100 m for cases
that required particularly large and computationally expensive domains.

The vertical domain spanned 17 km and included a
stretched grid spacing. The highest resolution for the stretched grid
was placed at the center of the initial cloud with grid size of 30
m. The vertical resolution decreased logarithmically to a maximum
grid spacing of approximately 300 m at the top of the model and approximately
400 m at the surface. A sponge layer was placed above 14 km to dampen
vertical motions at the top of the model and to prevent reflection
of gravity waves off the top of the model domain. The model time step
for dynamics was between 1.0 and 10.0 s and was chosen to be the largest
time step that was computationally stable.

For radiative transfer, the UU LESM model uses a
plane parallel broadband approach, using a $\delta$-four stream scheme
for parameterization of radiative transfer \citep{Liou88}, based
on the correlated k-distribution method \citep{FuLiou92}. Radiative
transfer calculations were performed at a time step of 60 s. Only
thermal radiation was considered in this study.

\begin{figure}[t]
\begin{centering}
\includegraphics[width=3in,angle=0]{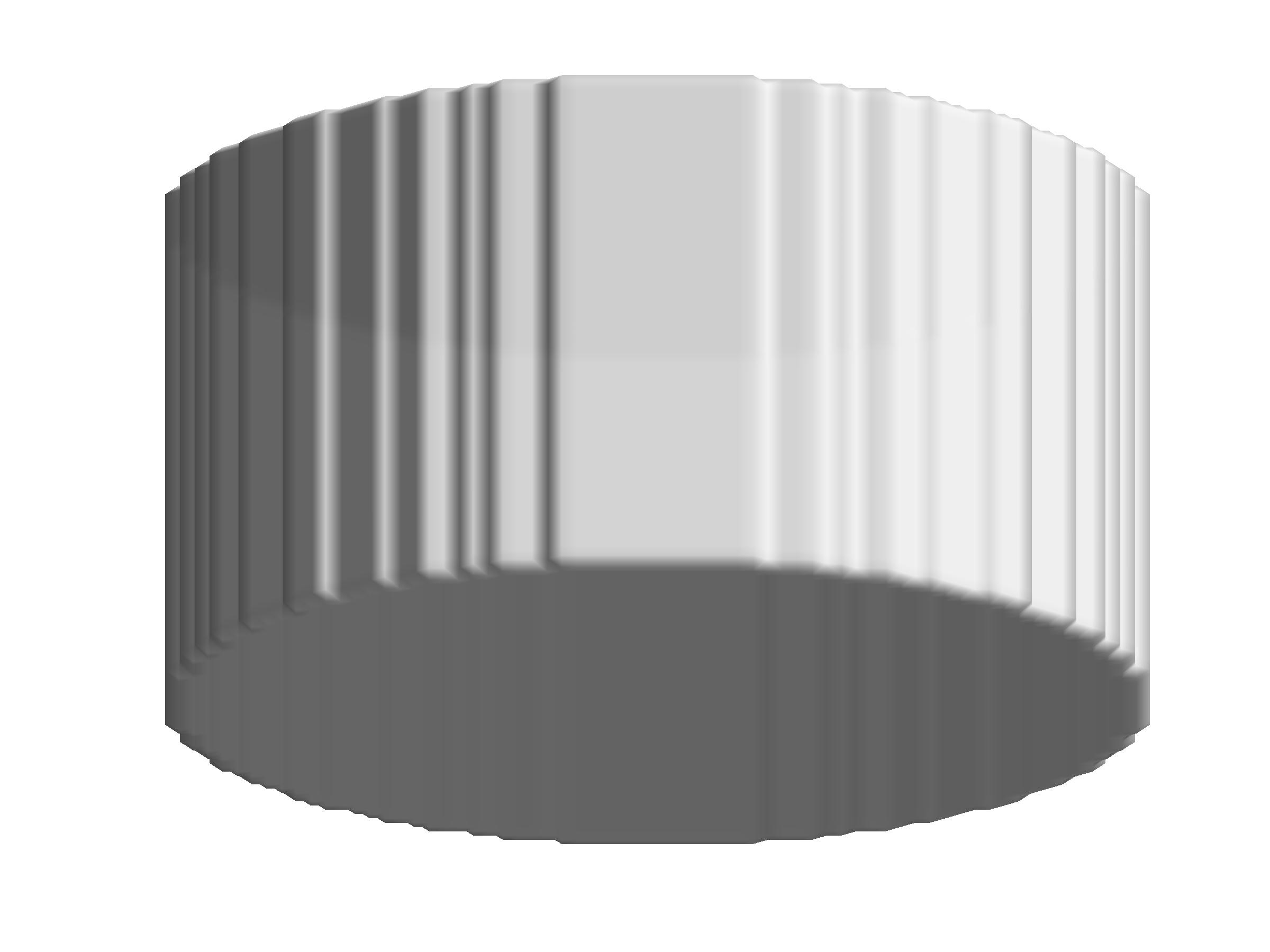}
\par\end{centering}

\caption{\label{fig:3D-initial}3D surface of cloud as initialized for all
simulations
}%
\end{figure}

For all cases examined, the model was initialized
with a standard tropical profile of temperature and atmospheric gases
with a buoyancy frequency $N$ of approximately 0.01 s$^{\text{-1}}$
throughout the depth of the model domain. Relative humidity was set
in two independent layers. In the bottom layer of the model, which
extends from the surface to 7.8 km, the relative humidity was set
to a constant 70\% with respect to liquid water. In the upper layer
of the model, from 7.8 km upwards, which contained the cloud between
8.8 km and 11.3 km, relative humidity with respect to ice was set
to a constant value of 70\%. All clouds were initialized as homogeneous
cylindrical ice clouds, as shown in Figure \ref{fig:3D-initial}.
Ice particles within the cloud were of uniform size with a fixed effective
radius of 20 $\mu$m and an initially uniform mixing ratio as prescribed
by the particular case. Cloud radius was prescribed according to the
particular case, but in each case the thickness was set to 2500 m
with the cloud base set at 8.8 km. Cloud base was chosen such that
the cloud top would be placed at approximately 200 mb, in rough accordance
with the average cirrus anvil height indicated by the Fixed Anvil
Temperature hypothesis \citep{HartmannAndLarson}. Both the cloud
and surrounding atmosphere were initialized to be at rest. No precipitation
was allowed in any of the model simulations. Cloud particle fall speed
was also neglected. All cases were run for one hour of model simulation
time.

Two cloud parameters were varied through several orders of magnitude
in order to explore a wide parameter space of possible evolutionary
behaviors. Cloud radius $L$ was chosen to be 100 m, 1 km, or 10 km.
The ice-water mixing ratio $q_{i}$ was set to 0.01, 0.1, or 1 g kg$^{-1}$.
This provided 9 unique combinations of cloud size and density, as
described in Tables \ref{tab:dimensionless-number-S} and \ref{tab:dimensionless-number-E},
that spanned a range of values of $S$ and $E$, and included combinations
that are sufficiently unstable that they are not observed naturally. 

\begin{table}[t]
\begin{centering}
\caption{\label{tab:dimensionless-number-S}Spreading number $S=\frac{\alpha_{\delta z}}{\alpha_{L}}=\frac{\mathcal{H}gL}{\theta N^{3}h^{2}}$}

\par\end{centering}

\centering{}\begin{tabular}{cccc}
\hline 
 & $L$=100m & 1km & 10km\tabularnewline
\hline 
$q_{i}$=0.01g/kg & 1.1$\times$10$^{-4}$ & 1.1$\times$10$^{-3}$ & 0.011\tabularnewline
0.1g/kg & 3.3$\times$10$^{-3}$ & 0.033 & 0.33\tabularnewline
1g/kg & 13 & 130 & 1300\tabularnewline
\hline
\end{tabular}%
\end{table}

\begin{table}[t]
\begin{centering}
\caption{\label{tab:dimensionless-number-E}Evaporation number $E=\frac{\alpha_{evap}}{\alpha_{\delta z}}=\frac{c_{p}\theta N^{2}h}{gL_{s}q_{i}}$}

\par\end{centering}

\centering{}\begin{tabular}{cccc}
\hline 
 & $L$=100m & 1km & 10km\tabularnewline
\hline 
$q_{i}$=0.01g/kg & 150 & 150 & 150\tabularnewline
0.1g/kg & 3.7 & 3.7 & 3.7\tabularnewline
1g/kg & 0.037 & 0.037 & 0.037\tabularnewline
\hline
\end{tabular}%
\end{table}

\begin{figure}[t]
\begin{centering}
\includegraphics[width=0.38\paperwidth,angle=0]{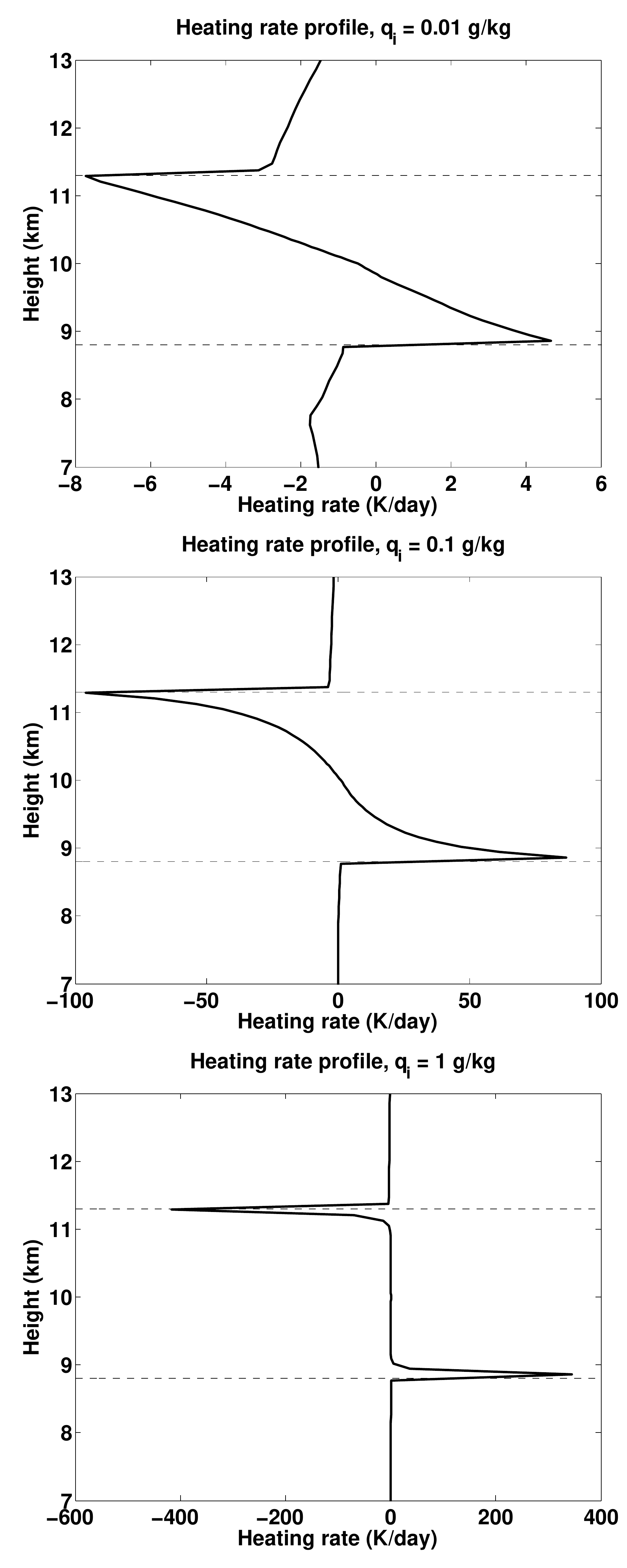}
\par\end{centering}

\centering{}\caption{\label{fig:Heating-rate-profile}Heating rate profiles for $q_{i}=0.01$
g kg$^{-1}$, $q_{i}=0.1$ g kg$^{-1}$, and $q_{i}=1$ g kg$^{-1}$.
Cloud vertical boundaries are marked with a dashed line.}
\end{figure}

Figure \ref{fig:Heating-rate-profile} shows the initial heating rate
profiles for each value of $q_{i}$ used in this study calculated
using the \citet{FuLiou92} radiative transfer parameterization. Note
that the heating is confined to a narrower layer at cloud base as
the ice water mixing ratio increases (Eq. \ref{eq:penetration depth}).
The heating profiles for both the $q_{i}=0.01$ g kg$^{-1}$ and $q_{i}=0.1$
g kg$^{-1}$ cases closely match the calculated heating rate profiles
from \citet{Lilly88}. However, the heating rate profile for the $q_{i}=1$
g kg$^{-1}$ case, which Lilly did not model, shows an order of magnitude
increase in the heating and cooling rates to several hundred K day$^{-1}$,
confined almost exclusively to the top and bottom of the cloud, with
virtually no heating in the interior.

For cases with $q_{i}=0.01$ g kg$^{-1}$, the radiative penetration
depth $h$ is 3300 m, which is deeper then the 2500 m cloud depth.
However, the heating profile is nearly linear through the depth of
the cloud with heating at cloud base and cooling at cloud top. Thus,
in cases where $q_{i}$ is 0.01 g kg$^{-1}$, the radiative penetration
depth $h$ is assumed to be half the cloud depth, or 1250 m, for the
purposes of calculating $S$ and $E$.

\section{Results}

In the parameter space of $S$ and $E$ described by Tables \ref{tab:dimensionless-number-S}
and \ref{tab:dimensionless-number-E}, tenuous and narrow clouds with
low values of ice water mixing ratio $q_{i}$ and cloud width $L$
have values of the spreading number $S$ that are less than 1. Theoretically,
such clouds are expected to undergo laminar lifting and spreading.
Tenuous clouds with large values of $E$ and small value of $S$ are
expected to evaporate at cloud base. Optically dense and broad clouds
with large values of $q_{i}$ and $L$ have values of $S$ much larger
than 1, and are expected to favor the concentration of potential energy
density in a thin layer at cloud base, leading to turbulent mixing
and erosion of stratified air within the cloudy interior. 

In what follows, numerical simulations are performed to test the validity
of the dimensionless numbers $S$ and $E$ for predicting cloud evolution.
Cases that describe the parameter space in $S$ will be discussed
first, since values of $E$ are relevant only for scenarios with $S<1$
where mixed-layer development is not the primary response to local
diabatic radiative heating.

\subsection{\label{sec:Isentropic-Adjustment}Isentropic Adjustment}

Simulations of clouds with values of $S<1$ are expected to show cross-isentropic
ascent of cloud base in response to local diabatic radiative heating
and, through continuity, laminar spreading. Effectively, the loss
of potential energy out the sides of the cloud (due to material flows)
is sufficiently rapid to maintain nearly flat isentropic surfaces
within the original cloud volume. Equivalently, cross-isentropic ascent
is sufficiently slow that the consequent horizontal pressure gradients
can be equilibrated through laminar spreading while keeping isentropic
surfaces approximately flat (Eq. \ref{eq:alpha-L}).

\begin{figure}[t]
\begin{centering}
\includegraphics[width=0.49\paperwidth,angle=0]{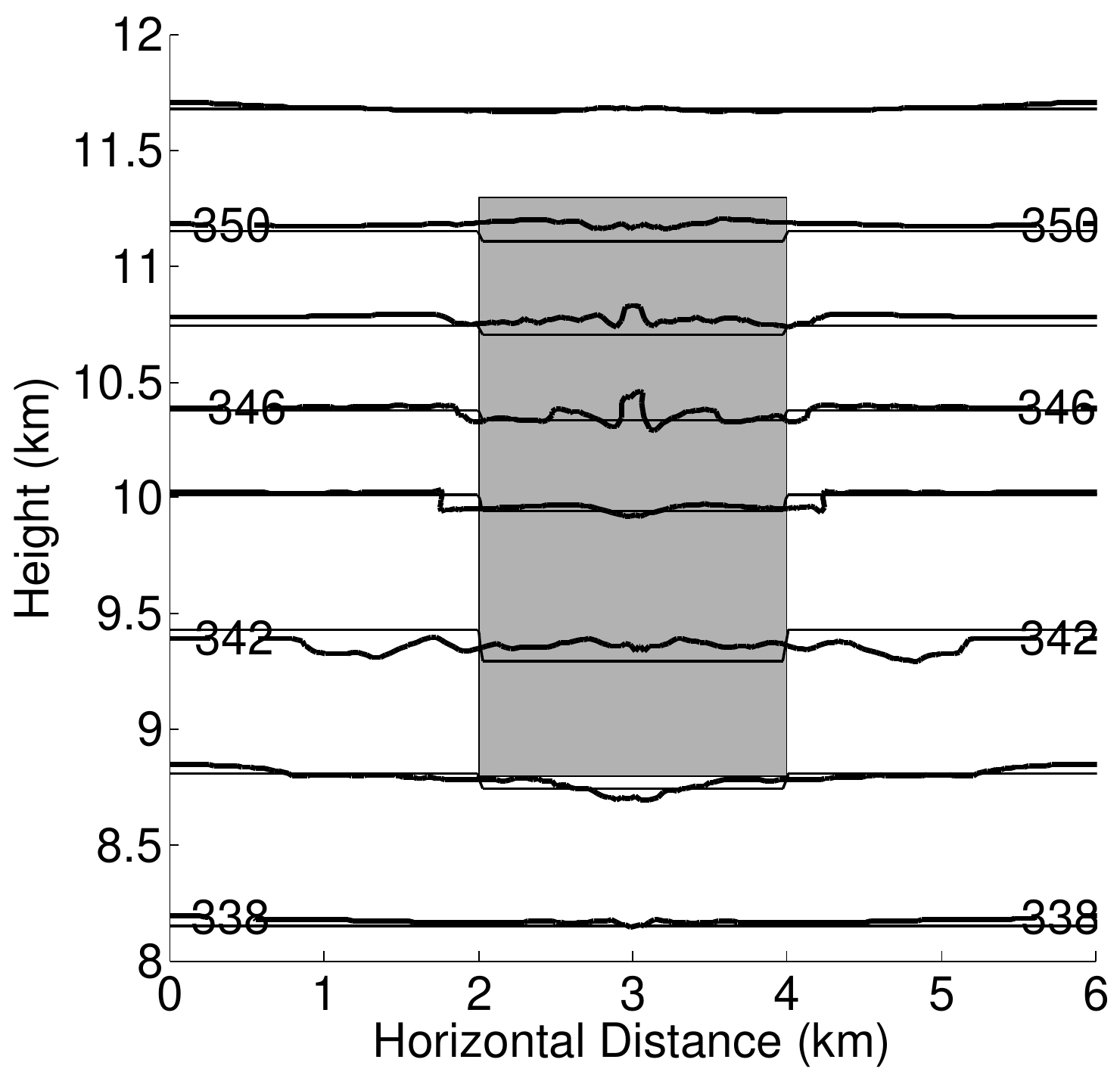}
\par\end{centering}

\caption{\label{fig:isentropic-adjustment-theta_e_cs}Cross section of $\theta_{e}$
contours through a cloud with $L$=1km and $q_{i}$=0.1 g kg$^{-1}$
($S=0.033$) after 0 s (thin) and 3600 s (thick) of simulation. The
initial cloud boundaries are indicated by the shaded region.}
\end{figure}

\begin{figure}[t]
\begin{centering}
\includegraphics[width=0.6\paperwidth,angle=0]{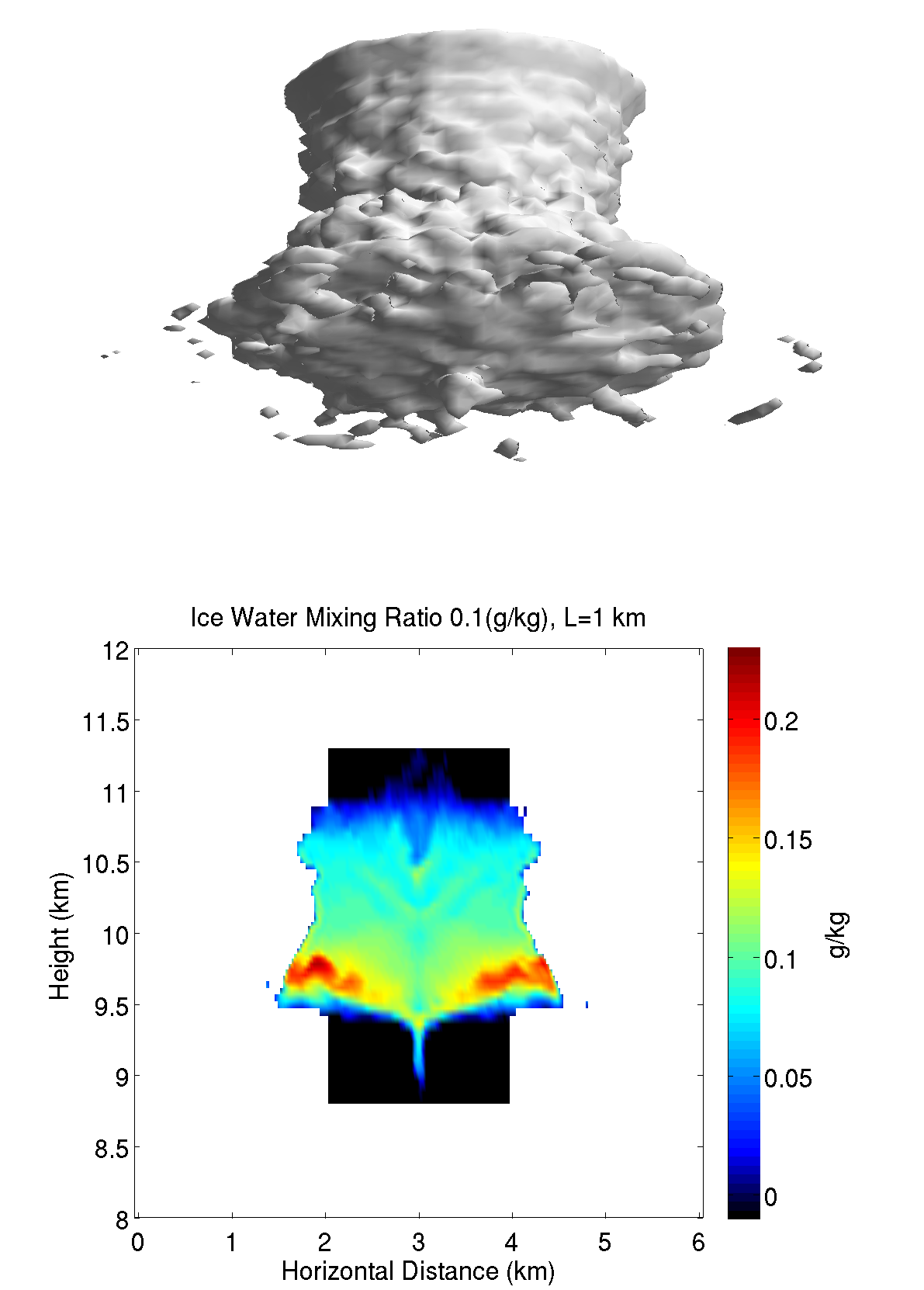}
\par\end{centering}

\caption{\label{fig:isentropic-adjustment-qi} as in Figure \ref{fig:isentropic-adjustment-theta_e_cs},
but a 3D surface plot of $q_{i}$ after 3600 s of simulation (top),
and a plot of a $q_{i}$ cross section of the cloud (bottom). The
initial position of the cloud is shown in black, while the state of
the cloud after 3600 s is shown in color, with the value of $q_{i}$
denoted by the color scale. Note the rise of cloud base and the horizontal
spreading.}
\end{figure}

A good example of this behavior is shown in a simulation of a cloud
with $L=1$ km and $q_{i}=0.1$ g kg$^{-1}$. This case has a value
of $S=0.033$, which implies that the primary response to radiative
heating should be adjustment through ascent across isentropic surfaces.
Figure \ref{fig:isentropic-adjustment-theta_e_cs} shows the isentropes,
or contours in $\theta_{e}$. The isentropes remain approximately
flat and unchanged from their initial state in response to the cross-isentropic
flow of cloudy air. As shown in Figure \ref{fig:isentropic-adjustment-qi},
the simulated cloud undergoes rising at cloud base and sinking at
cloud top, while spreading horizontally.

\subsection{\label{sec:Mixing}Mixing}

Clouds with values of $S>1$ are not expected to be associated with
laminar motions. Instead, radiative heating bends down isentropic
surfaces so rapidly as to create a local instability that cannot be
restored sufficiently rapidly by laminar cloud outflows (Eq. \ref{eq:alpha-delta-z}).
Radiative heating is sufficiently concentrated to initiate turbulent
mixing that produces a growing mixed-layer. Unlike the $S<1$ case,
isentropes do not stay flat. 

\begin{figure}[t]
\begin{centering}
\includegraphics[width=0.4\paperwidth,angle=0]{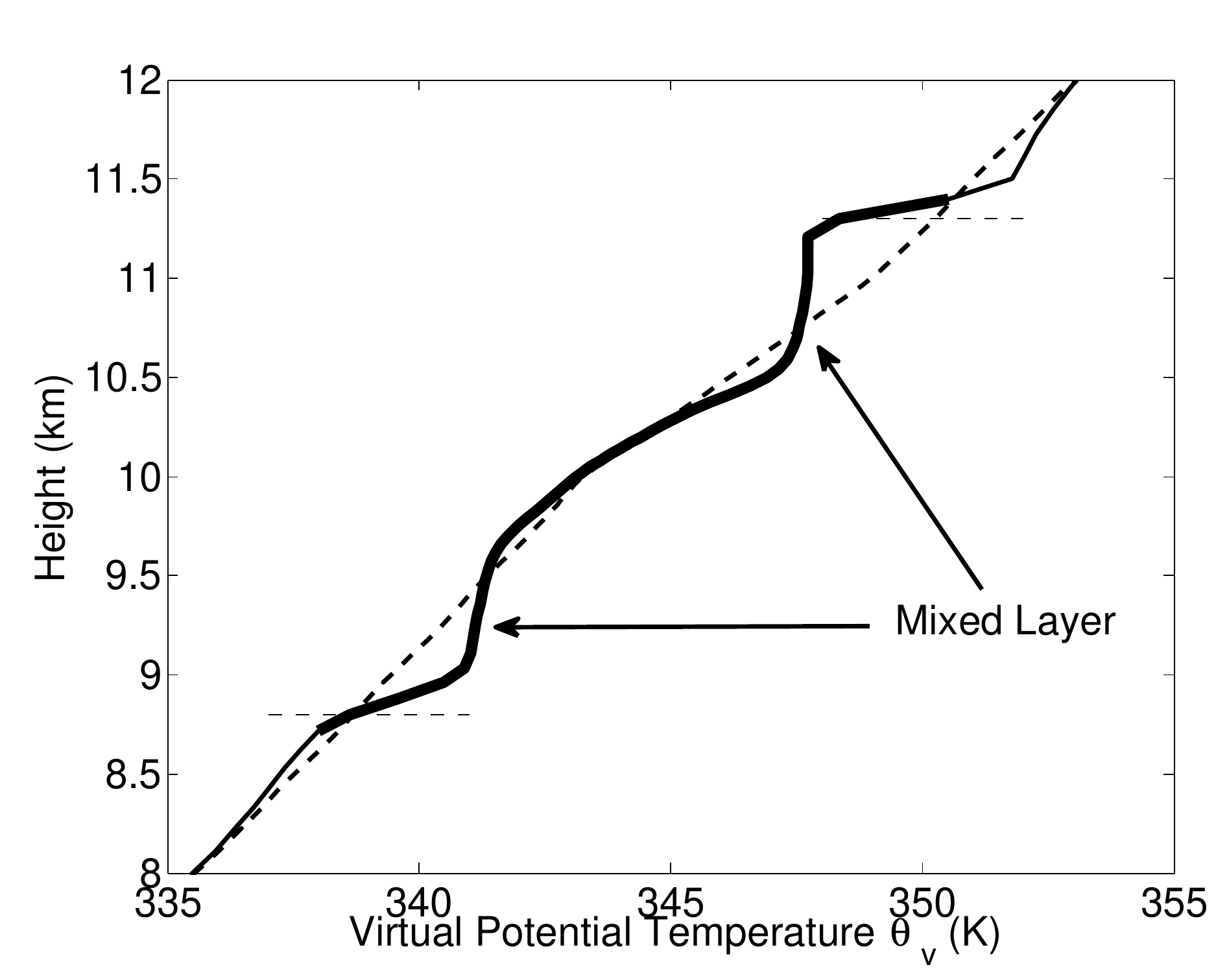}
\par\end{centering}

\caption{\label{fig:mixing-profile}$\theta_{v}$ profile of a cloud with $L$=10
km and $q_{i}$=1 g kg$^{-1}$ after 3600 s of simulation. The initial
profile is plotted in a dashed line with horizontal dashed lines indicating
initial cloud base and cloud top. The $\theta_{v}$ profile is calculated
as a horizontal average of all $\theta_{v}$ profiles within an annular
region of the cloud. The inner edge of the annulus is 7.5 km from
cloud center and the outer edge of the annulus is at 9 km from cloud
center.}
\end{figure}

An example, shown in Figure \ref{fig:mixing-profile}, is for a simulated
cloud that has initial condition values of cloud radius $L=10$ km
and ice water mixing ratio $q_{i}=1$ g kg$^{-1}$. Since $S=1300$,
it is expected that the potential energy density at cloud base will
increase at a rate that is faster than the loss rate of potential
energy through cloud lateral expansion (Eqs. \ref{eq:alpha-delta-z}
and \ref{eq:alpha-L}). A mixed layer will develop because the deposition
of radiative energy creates buoyancy that does work to overcome the
static stability of overlying cloudy air and create a mixed layer.
Meanwhile the mixed-layer expands with speed $u_{mix}=N\delta z$
(Eq. \ref{eq:u_mix}), where $\delta z$ is the mixed-layer depth
and $N$ is the static stability of air surrounding the cloud.

\begin{figure}[t]
\begin{centering}
\includegraphics[width=0.49\paperwidth,angle=0]{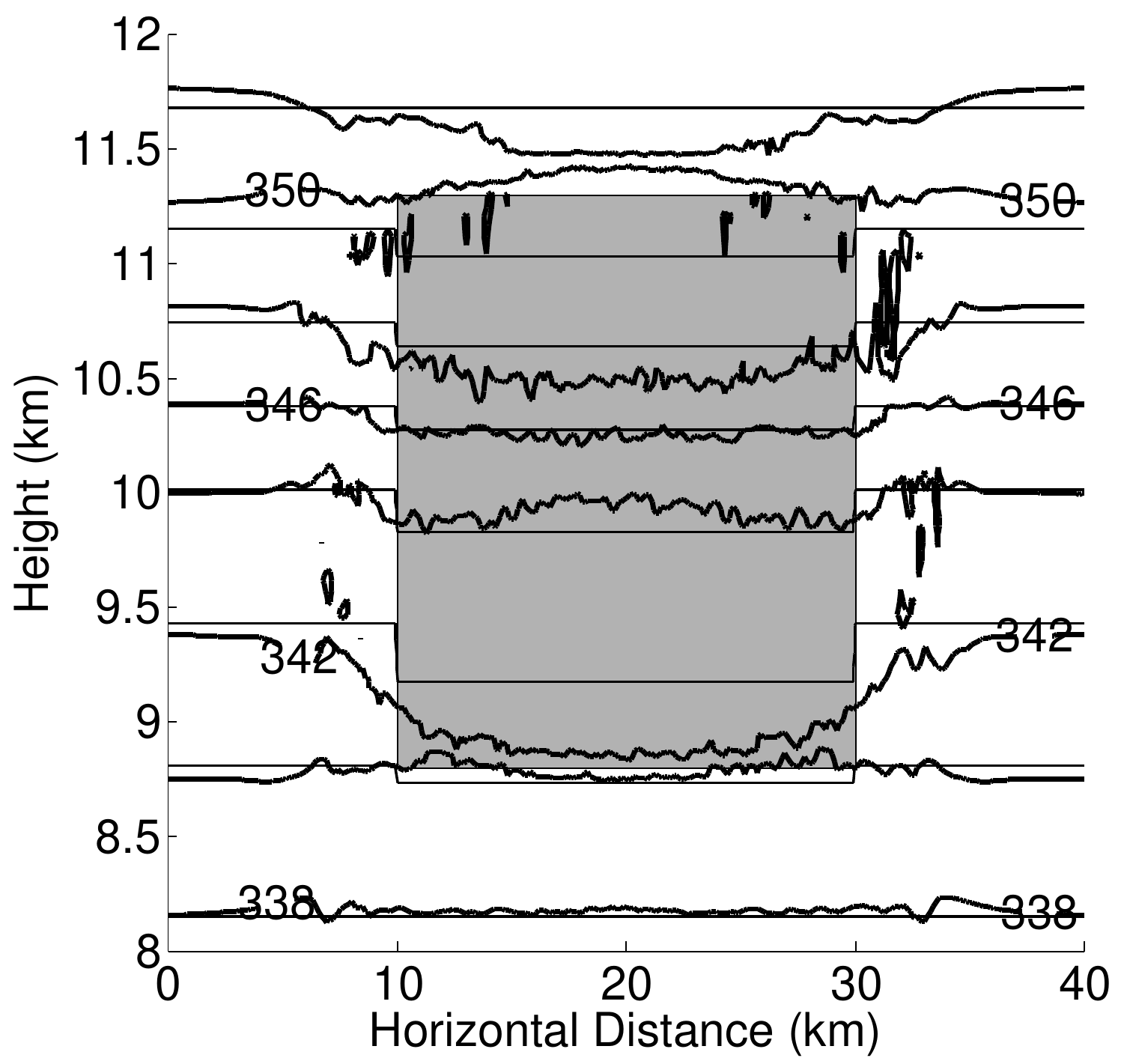}
\par\end{centering}

\caption{\label{fig:mixing-theta-e-cs}As in Figure \ref{fig:mixing-profile}
but a cross section of $\theta_{e}$ contours through a cloud after
0 s (thin) and 3600 s (thick) of simulation. The initial cloud boundaries
are indicated by the shaded region.}
\end{figure}

The numerical simulations reproduce these features. A mixed-layer
can be seen in the $\theta_{v}$ profile plotted in Figure \ref{fig:mixing-profile},
showing the average cloud properties after 1 hour of model simulation.
This profile is a horizontally averaged profile taken within 9 km
of cloud center. On average, the mixed-layer exhibits a nearly adiabatic
profile in $\theta_{v}$. At 1 h simulation time, the mixed-layer
at cloud base is nearly 800 m deep. The mixed layer expands horizontally
\textit{along} isentropes, as seen in Figure \ref{fig:mixing-theta-e-cs}.
The {}``bowl'' shaped spreading of the cloud is because intense
radiative heating at cloud base bends isentropic surfaces downward. 

\begin{figure}[t]
\begin{centering}
\includegraphics[width=0.6\paperwidth,angle=0]{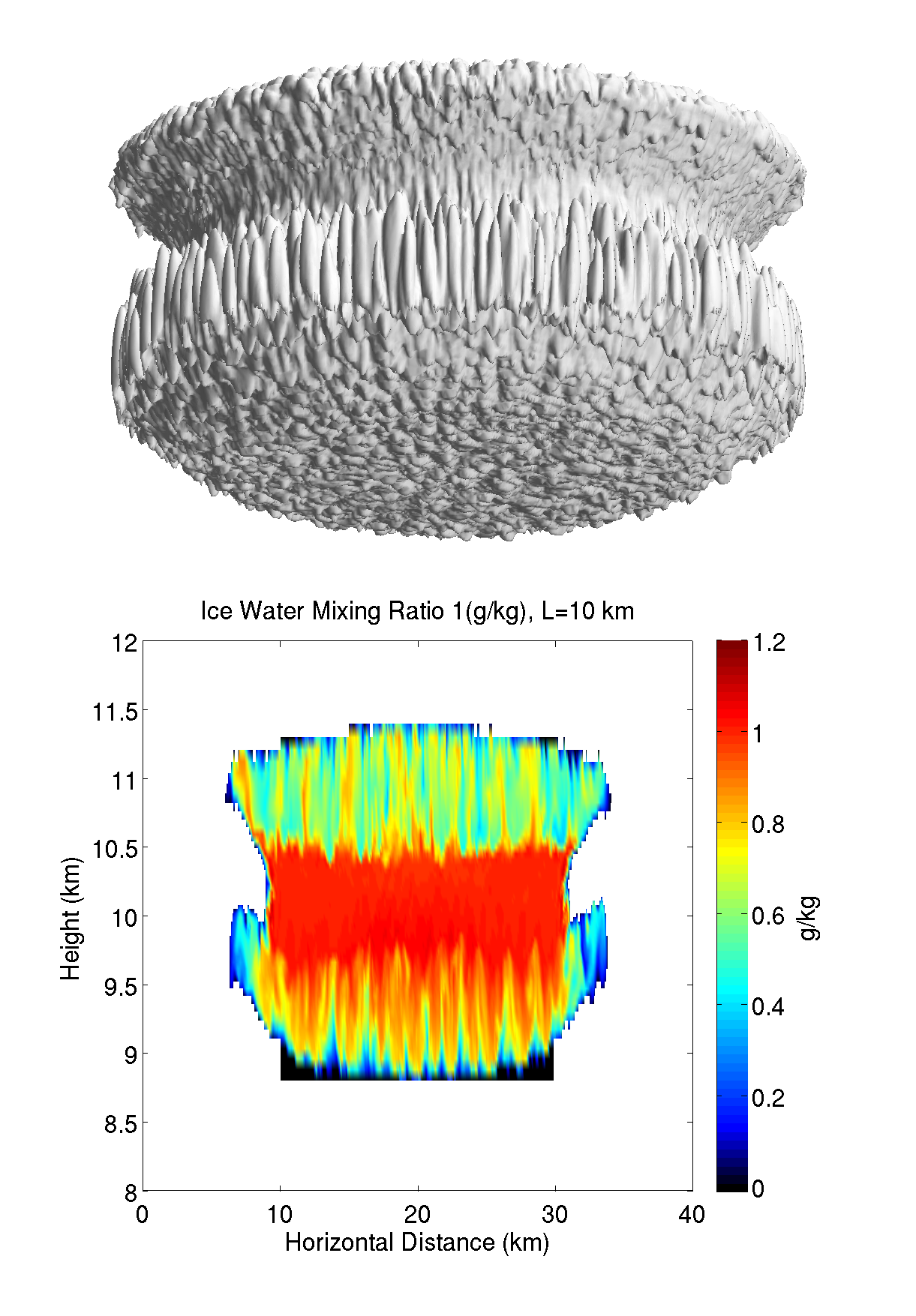}
\par\end{centering}

\caption{\label{fig:mixing-qi} As in Figure \ref{fig:mixing-profile}, but
a 3D surface plot of $q_{i}$ after 3600 s of simulation (top), and
a plot of a $q_{i}$ cross section of the cloud (bottom). The initial
position of the cloud is shown in black, while the state of the cloud
after 3600 s is shown in color, with the value of $q_{i}$ denoted
by the color scale. Note that cloud base remains at roughly the same
level and that the cloud bends upward as it spreads outward. }
\end{figure}

This mixed-layer development and spreading can also be seen in cross
sectional plots of $q_{i}$ in Figure \ref{fig:mixing-qi}. There
is a mixed-layer at both cloud base and cloud top with darker shading
indicating where drier air has been entrained from below or above.
Note that cloud base and cloud top remain at roughly constant elevation.
In Figure \ref{fig:isentropic-adjustment-theta_e_cs}, for a case
where $S\ll1$, radiative flux convergence at cloud base drives cross-isentropic
laminar ascent. In this case, where $S\gg1$, laminar ascent does
not occur. Instead, cloud base remains nearly at its initial vertical
level and there is formation of a turbulent mixed layer that spreads
outward along isentropes. Notably, the mixed-layer circulations at
cloud base have a mammatus-like quality to them, something we have
discussed more extensively in \citet{Mammatus}.

These behaviors can be quantified by examination of  the rapidity of development
of a well-mixed layer at cloud base. If the dominant mode of evolution
is cross-isentropic lofting, then vertical potential temperature gradients
should remain relatively undisturbed. Conversely, if mixing is the
dominant response, then potential temperature will evolve to become
more constant with height. 

Table \ref{tab:disturbance} shows the cloud domain-averaged, logarithmic
rate of decrease in the static stability $d\ln N^{2}/dt$, where $N^{2}\propto d\theta_{v}/dz$.
Calculations are evaluated for the lowermost 80 m of the cloud within
the initial 360 s of simulation time. The destabilization of cloud
base reflects the magnitude of the Spreading Number $S$ (Table \ref{tab:dimensionless-number-S}),
with large values of $S$ demonstrating the most rapid rates of mixed-layer
development. 

\begin{table}[t]
\caption{\label{tab:disturbance}Rate of destabilization at cloud base ($-d\ln N^{2}/dt$)
in units of hr$^{-1}$. Cases with a Spreading Number $S$ that is
much greater than one are indicated in bold.}

\centering{}\begin{tabular}{cccc}
\hline 
 & $L$=100m & 1km & 10km\tabularnewline
\hline 
$q_{i}$=0.01g/kg & 0.12 & 0.17 & 0.15\tabularnewline
0.1g/kg & 1.32 & 0.83 & 1.94\tabularnewline
1g/kg & \textbf{4.00} & \textbf{11.42} & \textbf{23.88}\tabularnewline
\hline
\end{tabular}%
\end{table}

\subsection{\label{sec:Evaporation}Evaporation}

Cloud bases with $S<1$ and $E>1$ are expected to evaporate more
quickly than they loft across isentropes (Table \ref{tab:dimensionless-number-E}).
For example, for a cloud with $L=1$ km and $q_{i}=0.01$ g kg$^{-1}$
, the calculated value of the Spreading Number $S$ is 0.0011, and
the value of the Evaporation number $E$ is 150. Based on these values,
the expected evolution of cloud base would be gradual evaporative erosion
of cloud base. 

To quantify the importance of evaporation to cloud evolution, the
rate of change in cloud mass $d\ln m/dt$, where $m$ is the mass
of cloud ice, was calculated over the first 180 s of simulation, but
only within the lower layer in which radiation from the surface is
absorbed, $h$, rather than the entire cloud. The absorptive layers
were taken to be 30 m, 300 m, and 1250 m for cloud ice water mixing
ratios of 1 g kg$^{-1}$, 0.1 g kg$^{-1}$, and 0.01 g kg$^{-1}$
respectively. 

\begin{table}[t]
\caption{Evaporation rate, in units of h$^{-1}$, defined here as the negative
of the logarithmic rate of mass change in the lower depth $h$ of
the cloud nto the cloud within the initial 180 s of simulation in
units of s for cases where $E>1$ and $S<1$\label{tab:Log-mass-change}}

\centering{}\begin{tabular}{ccc}
\hline 
 & $E$ = 150 & $E$ = 3.7\tabularnewline
\hline 
$L$ = 100 m & 5.8 & 0.79\tabularnewline
1 km & 4.0 & 0.72\tabularnewline
10 km & 1.5 & 0.68\tabularnewline
\hline
\end{tabular}%
\end{table}

From Eq. \ref{eq:alpha_evap} for $\alpha_{evap}$, the anticipated
evaporation rate at cloud base is approximately $7$ h$^{-1}$ based
on the modeled net flux absorption $\Delta F$ of 74 $W$ m$^{-2}$
within the absorption layer $h$. Table \ref{tab:Log-mass-change}
shows maximum modeled evaporation rates that are nearly as large,
at least where $E$ is maximized and the cloud is narrow. However,
rates of evaporation decrease with increasing cloud width $L$, perhaps
because $S$ increases and stronger dynamic motions at cloud base
replace evaporated cloud condensate with newly formed cloud matter.
In general, however, tenuous cirrus clouds are most susceptible to
erosion by evaporation at cloud base, particularly if they are not
very broad.

\subsection{Precipitation}

While the role of precipitation has been excluded from these simulations
in order to clarify the physical behavior, certainly natural clouds
can have significant precipitation rates. An estimate of the relative
importance of precipitation is briefly discussed here.

The characteristic precipitation timescale $\alpha_{precip}$ depends
on the rate of depletion of cloud water by precipitation $P$ and
the average ice water content $IWC$. For example, in a cirrus anvil
in Florida measured by aircraft during the CRYSTAL-FACE field campaign,
the measured value of $P$ was 0.05 g m$^{-3}$ h$^{-1}$ compared
to values of $IWC$ of 0.3 g m$^{-3}$ \citep{GarrettEtAl2005}, implying
a precipitation depletion rate $\alpha_{precip}=P/IWC=0.15$ h$^{-1}$.
For comparison, corresponding values for the radiative adjustment
rates are $\alpha_{L}\simeq0.11$ h$^{-1}$ (Eq. \ref{eq:alpha-L})
and $\alpha_{\delta z}\simeq144$ h$^{-1}$ (Eq. \ref{eq:alpha-delta-z}).
While development of a turbulent mixed layer is the fastest process,
precipitation depletes cloud condensate at a rate that is comparable
to $\alpha_{L}$, the rate at which gravitational equilibrium is restored
through cross-isentropic flows and laminar spreading.

\section{Discussion}

We have separated the evolutionary response of clouds to local diabatic
heating into distinct modes of cross-isentropic lifting, along-isentropic
spreading, and evaporation of cloud condensate. A straightforward
method has been described for determining how a cloud will evolve
based on ratios of the associated rates. The dominant modes of evolution
are outlined in Table \ref{tab:evolution mode-1}. 

\begin{table}[t]
\caption{\label{tab:evolution mode-1}Dominant modes of evolution observed
in simulations. Cases where the Spreading Number $S$ and Evaporation
Number $E$ are much greater than one are indicated in bold and italics,
respectively.}

\centering{}\begin{tabular}{cccc}
\hline 
 & $L$=100m & 1km & 10km\tabularnewline
\hline 
$q_{i}$=0.01g/kg & \emph{evaporation} & \emph{evaporation} & \emph{evaporation}\tabularnewline
0.1g/kg & lofting & lofting & mixing\tabularnewline
1g/kg & \textbf{mixing} & \textbf{mixing} & \textbf{mixing}\tabularnewline
\hline
\end{tabular}%
\end{table}

For example, cirrus anvils begin their life cycle as dense cloud from
convective towers that have reached their level of neutral buoyancy \citep{CloudNomenclature,JonesEtAl,TC4}.
Such broad optically thick clouds are associated with high values
of the spreading number $S$ due to their large horizontal extent
and high concentrations of cloud ice. Radiative flux convergence is
confined to a thin layer at cloud base. Heating is so intense, and
the cloud is so broad, that the cloudy heated air cannot easily escape
by spreading into surrounding clear air. Instead, large values of
$S$ favor the development of a deepening mixed-layer. The mixed-layer
still spreads, but in the form of turbulent density currents rather
than laminar motions. 

However, as the cloud spreads and thins, the value of the spreading
number $S$ evolves. $S$ is proportional to the heating rate $\mathcal{H}$,
cloud width $L$, and inversely proportional to the square of the
depth of the mixed layer $\delta z^{2}$ (Eq. \ref{eq:Spreading-Number}).
Cloud spreading increases the value of $L$, and this acts as a positive
feedback on $S$. But as the cloud spreads, the mixed-layer depth
increases as $t^{1/2}$ (Eq. \ref{eq:t-one-half}), progressively
diluting the impact of radiative heating on dynamic development by
a factor of $\delta z/h$. Thus, while cloud spreading increases $S$,
this is offset by increasingly diluted heating rates within the mixed-layer
(Eq. \ref{eq:Spreading-Number}).

From Eq. \ref{eq:Spreading-Number}, $S$ can be rewritten as \begin{equation}
S=A\frac{L}{\delta z^{3}}\end{equation}
where $A=\mathcal{H}gh/\theta_{v}N^{3}$ is assumed to be constant,
assuming here that $q_{i}$ is fixed. Thus, the rate of change in
$S$ is given by \begin{equation}
\frac{d\ln S}{dt}\mid_{q_{i}}=\frac{d\ln L}{dt}-3\frac{d\ln\delta z}{dt}\label{eq:dlnS-basic}\end{equation}
From Eq. \ref{eq:mix-layer-depth}, and since $dL/dt=u_{mix}\sim N\delta z$
(Eq. \ref{eq:u_mix}), Eq. \ref{eq:dlnS-basic} can be rewritten as
\begin{equation}
\frac{d\ln S}{dt}\mid_{q_{i}}=\frac{N\delta z}{L_{0}}-\frac{3NA}{\delta z^{2}}\label{eq:dlnS-explicit}\end{equation}
Finally, from Eq. \ref{eq:t-one-half}, if the mixed layer depth evolves
over time as $\delta z=(NAt)^{1/2}$, Eq. \ref{eq:dlnS-explicit}
becomes \begin{equation}
\frac{d\ln S}{dt}=\frac{(N^{3}At)^{1/2}}{L_{0}}-\frac{3}{t}\label{eq:time-balance}\end{equation}

Thus, the evolution of $S$ is controlled by two terms, the first
being a positive feedback related to cloud spreading, and the second
being a negative feedback related to mixed-layer deepening. Provided
that\begin{equation}
t<t_{max}=\left(\frac{9L_{o}^{2}}{AN^{3}}\right)^{1/3}\end{equation}
the negative feedback dominates, so that to a good approximation \begin{equation}
\frac{d\ln S}{d\ln t}\simeq-3\end{equation}
which can be solved for the general solution \begin{equation}
S(t)\simeq S_{o}\left(\frac{t_{0}}{t}\right)^{3}\label{eq:Soft}\end{equation}

For a thick cirrus anvil with initial values of $q_{i}$ of 1 g kg$^{-1}$,
$L$ of 10 km, and $S$ of 1300, the value of $A$ is 3510 m$^{2}$
and $t_{max}\simeq10$ h. By comparison, from Eq. \ref{eq:Soft},
the value of $S$ rapidly drops to a value of approximately unity
within time $t\simeq10t_{0}$. While the value of $t_{0}$ is not
explicitly defined, assuming that it is one buoyancy period $2\pi/N$,
then the time scale for the cirrus anvil to shift from turbulent mixing
to isentropic adjustment is of order one hour. Because this timescale
is much less than $t_{max}$, the anvil never manages to enter a regime
of runaway mixed-layer deepening where Eq. \ref{eq:time-balance}
is positive. What is interesting is that this timescale for a convecting
anvil to move into a laminar flow regime is comparable to the few
hours lifetime of tropical cirrus associated with deep-convective
cloud systems \citep{cloudLifetime}. A transition to laminar behavior
seems inevitable.

\begin{figure}[t]
\begin{centering}
\includegraphics[width=0.6\paperwidth,angle=0]{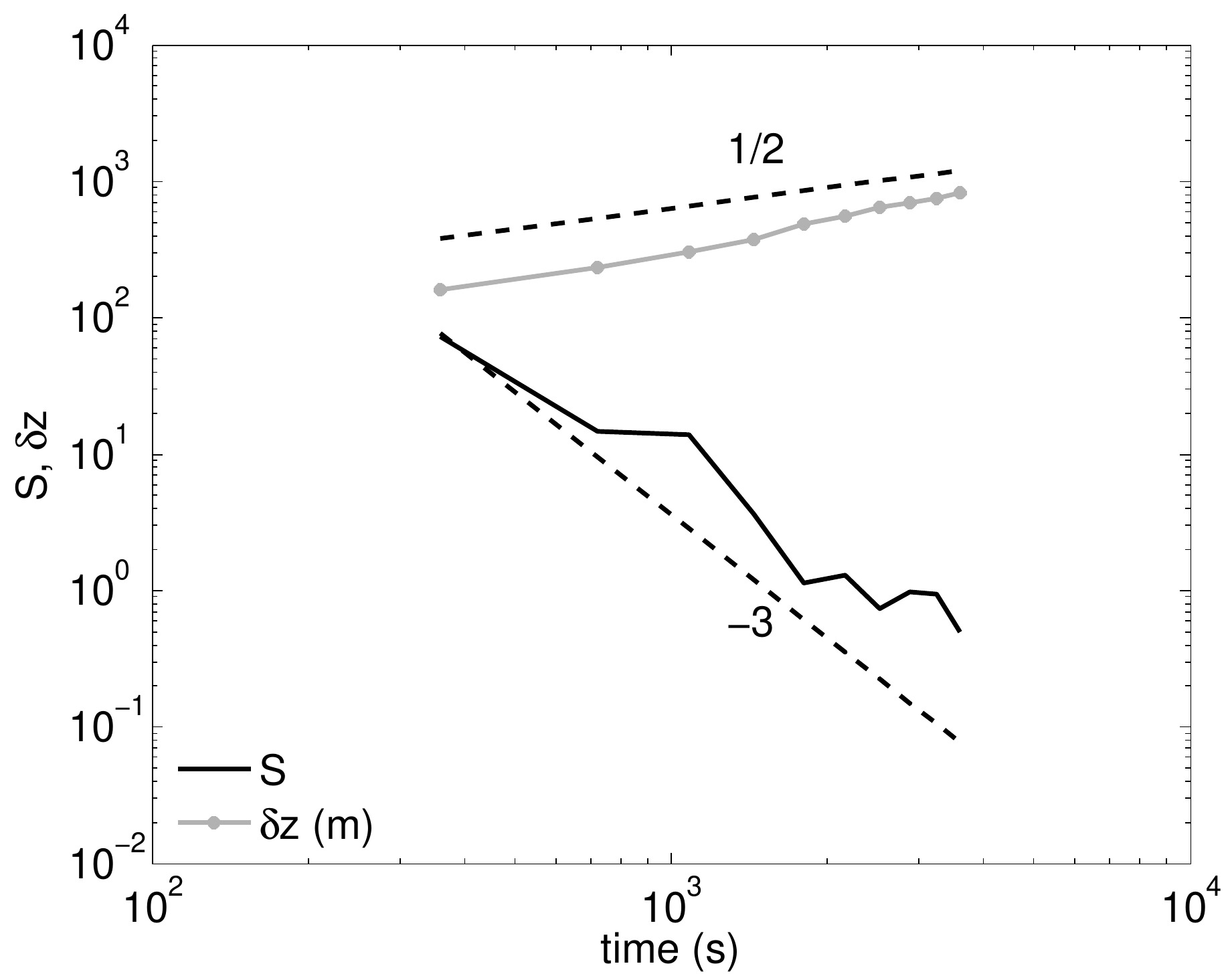}
\par\end{centering}

\caption{\label{fig:S(t)}The time evolution of the Spreading Number $S$ and
the mixed-layer depth $\delta z$ of a cloud with ice water mixint
ratio $q_{i}$ of 1 g kg$^{-1}$ and width $L$ of 10 km from 0 to
3600 s. The dashed lines indicates slopes of 1/2 and -3 on the log-log plot as indicated.}
\end{figure}

Figure \ref{fig:S(t)} shows numerical simulations for the time evolution
of $S$ within the cloud base domain. These reproduce the theoretically
anticipated decay at a rate close to the anticipated $t^{-3}$ power
law. The decay in $S$ is dominated by mixed-layer deepening, which
roughly follows the anticipated $t^{1/2}$ power law. It may seem
counter-intuitive, but it is deepening of a turbulent mixed-layer
that allows for a transition to laminar behavior: current radiative
flux deposition becomes increasingly diluted in past deposition. Once
an anvil reaches $S\sim1$, the rate at which the mixed layer deepens
becomes roughly equal to the rate at which laminar flow restores gravitational
equilibrium through spreading. At this point, the dynamic evolution
of the cirrus anvil enters a new regime where it adjusts to any radiatively
induced gravitational disequilibrium through either cross-isentropic
lofting \citep{Danielsen82,Ackerman88} or evaporation \citep{JensenEtAl96}. 

As a contrasting example, contrail formations are typically optically
thin and horizontally narrow. In some cases they can evolve into broad
swaths of cirrus that persist for up to 17 hours after initial formation
and radiatively warm the surface \citep{contrailClimate}. While we
did not specifically model contrails in this study, the theoretical
principles that we discussed can provide guidance for how they might
be expected to evolve. 

Immediately following ejection from a jet engine, the contrail air
has water contents of a few tenths of a gram per meter cubed \citep{contrailDensity},
contained within a very narrow horizontal domain \citep{CirrusSize}.
In this case, the cloud can be characterized in an idealized sense
by $q_{i}=$1 g kg$^{-1}$ and $L=$100 m (Table \ref{tab:dimensionless-number-S}).
Since the expressions for $A$ and $t_{max}$ discussed above do not
depend on the horizontal extent of the cloud $L$, their values are
identical to those of the idealized anvil that was explored. However,
the initial value of $S$ does depend on $L$, and with an initial value of 13 it is one hundred
times smaller than for the anvil case. Since
the initial value for $S$ is still larger than unity, it should be
expected that the contrail cirrus will be able to sustain radiatively
driven turbulent mixing in its initial stages. However, from Eq.
\ref{eq:Soft}, $S$ should be expected to decline to unity in about
20 minutes, at which point more laminar circulations take over that
allow for the contrail cloud to spread laterally while slowly lofting
across isentropes.

\section{Conclusions}

In this study, the evolutionary behavior of idealized clouds in response
to local diabatic heating was estimated from simple theoretical arguments
and then compared to high resolution numerical simulations. Simulated
clouds were found to evolve in a manner that was consistent with expected
behaviors. Dense, broad clouds had high initial values of a spreading
number $S$ (Eq. \ref{eq:Spreading-Number}) and formed deepening
convective mixed-layers at cloud base that spread in turbulent density
currents. The mixed-layers were created because isentropic surfaces
were bent downward by radiative flux convergence to create a layer
of instability. The mixed-layer deepened at a rate $\alpha_{\delta z}$
(Eq. \ref{eq:alpha-delta-z}) that was much faster than the rate at
which the potential instability could be restored through along-isentropic
outflow into surrounding clear air at rate $\alpha_{L}$ (Eq. \ref{eq:alpha-L}).
For particularly high values of $S$, the mixed-layer production from
radiative heating was so strong as to create mammatus clouds at cloud
base \citep{Mammatus}.

Tenuous and narrow clouds with initial values of $S<1$ displayed
gradual laminar ascent of cloud base across isentropic surfaces while
the cloud spread through continuity into surrounding clear sky. Isentropic
surfaces stayed roughly flat because the rate of along-isentropic
spreading $\alpha_{L}$ (Eq. \ref{eq:alpha-L}) was sufficiently rapid
compared to the rate of cross-isentropic lifting $\alpha_{\delta z}$
(Eq. \ref{eq:alpha-delta-z}) that isentropic surfaces in the cloud
were continuously returned to their original equilibrium heights.
Clouds with low values of $S$ and also high values of an evaporation
number $E$ (Eq. \ref{eq:Evaporation Number}) tended to evaporate
quickly because the rate at which cloud condensate evaporated $\alpha_{evap}$
(Eq. \ref{eq:alpha_evap}) was much faster than the rate at which
the cloud layer rose in cross-isentropic laminar ascent $\alpha_{L}$
(Eq. \ref{eq:alpha-L}).

For clouds with values of $S$ that are initially high, the tendency
is that $S$ falls with time as the convergence of radiative flows
at cloud base becomes increasingly diluted in a deepening mixed-layer.
We found that dense cirrus anvils with a large horizontal extent remain
in a mixed-layer deepening regime for nearly an hour before shifting
across the $S=1$ threshold into a cross-isentropic laminar lofting
regime. Contrail cirrus are expected to make the same transition,
but in a matter of tens of minutes.

It is important to note that the precision of any of these results
is limited by the simplifications that were taken. Most important
is that no precipitation was included in the numerical simulations,
so simulated clouds presumably persisted longer than if precipitation
were included. Also, single-sized ice particles were used rather than
a distribution of ice particle sizes. Gravitational sorting would
result in a higher concentration of larger ice particles near cloud
base and a higher concentration of small ice particles near cloud
top \citep{GarrettEtAl2005,Jensen2010TTL}. 

Nonetheless, it has been shown that local diabatic heating heating
can drive dynamic motions and microphysical changes that are at least
as important as precipitation, and easily predicted from the simple
calculation of two dimensionless numbers. A practical future application
of this work might be improved constraints of the fast, smale-scale
evolution of fractional cloud coverage within a GCM gridbox, limiting
the need for explicit, and expensive, fluid simulations of sub-grid
scale processes. 

%

\ifthenelse{\boolean{dc}}
{}
{\clearpage}
\bibliographystyle{ametsoc}
\bibliography{references}

\begin{thebibliography}{29}
\providecommand{\natexlab}[1]{#1}
\providecommand{\url}[1]{\texttt{#1}}
\providecommand{\urlprefix}{URL }
\expandafter\ifx\csname urlstyle\endcsname\relax
  \providecommand{\doi}[1]{doi:\discretionary{}{}{}#1}\else
  \providecommand{\doi}{doi:\discretionary{}{}{}\begingroup
  \urlstyle{rm}\Url}\fi
\providecommand{\eprint}[2][]{\url{#2}}

\bibitem[{Ackerman et~al.(1988)Ackerman, Liou, Valero, and
  Pfister}]{Ackerman88}
Ackerman, T.~P., K.~N. Liou, F.~P.~J. Valero, and L.~Pfister, 1988: Heating
  rates in tropical anvils. \textit{J. Atmos. Sci.}, \textbf{45}, 1606--1623.

\bibitem[{Burkhardt and Kärcher(2011)}]{contrailClimate}
Burkhardt, U. and B.~Kärcher, 2011: Global radiative forcing from contrail
  cirrus. \textit{Nature Climate Change}, \textbf{1}, 54--58,
  doi:10.1038/nclimate1068.

\bibitem[{Cole et~al.(2005)Cole, Barker, Randall, Khairoutdinov, and
  Clothiaux}]{cole2005}
Cole, J. N.~S., H.~W. Barker, D.~A. Randall, M.~F. Khairoutdinov, and E.~E.
  Clothiaux, 2005: Global consequences of interactions between clouds and
  radiation at scales unresolved by global climate models. \textit{Geophys.
  Res. Lett.}, \textbf{32}, L06\,703, doi:10.1029/2004GL020\,945.

\bibitem[{Danielsen(1982)}]{Danielsen82}
Danielsen, E.~F., 1982: A dehydration mechanism for the stratosphere.
  \textit{Geophys. Res. Lett.}, \textbf{9}, 605--608.

\bibitem[{Dinh et~al.(2010)Dinh, Durran, and Ackermann}]{Dinh&Durran10}
Dinh, T.~P., D.~R. Durran, and T.~Ackermann, 2010: The maintenance of tropical
  tropopause layer cirrus. \textit{J. Geophys. Res.}, \textbf{115}, D02\,104,
  doi:10.1029/2009/JD012\,735.

\bibitem[{Dobbie and Jonas(2001)}]{DobbieAndJonas2001}
Dobbie, S. and P.~Jonas, 2001: Radiative influences on the structure and
  lifetime of cirrus clouds. \textit{Q.J.R.Meteorol.Soc.}, \textbf{127},
  2663--2682, doi:10.1002/qj.49712757\,808.

\bibitem[{Dufresne and Bony(2008)}]{Dufresne&Bony08}
Dufresne, J.~L. and S.~Bony, 2008: An assessment of the primary sources of
  spread of global warming estimates from coupled ocean-atmosphere models.
  \textit{J. Climate}, \textbf{21}, 5135--5144, doi:10.1175/2008JCLI2239.1.

\bibitem[{Durran et~al.(2009)Durran, Dinh, Ammerman, and Ackerman}]{Durran09}
Durran, D.~R., T.~Dinh, M.~Ammerman, and T.~Ackerman, 2009: The mesoscale
  dynamics of thin tropical tropopause cirrus. \textit{J. Atmos. Sci},
  \textbf{66}, 2859--2873, doi:10.1175/2009JAS3046.1.

\bibitem[{Fu and Liou(1992)}]{FuLiou92}
Fu, Q. and K.~N. Liou, 1992: On the correlated k-distibution method for
  radiative transfer in nonhomogeneous atmospheres. \textit{J. Atmos. Sci.},
  \textbf{49}, 2139--2156.

\bibitem[{Garrett et~al.(2010)Garrett, Schmidt, Kihlgren, and
  Cornet}]{Mammatus}
Garrett, T.~J., C.~T. Schmidt, S.~Kihlgren, and C.~Cornet, 2010: Mammatus
  clouds as a response to cloud-base radiative heating. \textit{J. Atmos.
  Sci.}, \textbf{67}, 3891--3903, doi:10.1175/2010JAS3513.1.

\bibitem[{Garrett et~al.(2006)Garrett, Zulauf, and Krueger}]{Garrett06}
Garrett, T.~J., M.~A. Zulauf, and S.~K. Krueger, 2006: Effects of cirrus near
  the tropopause on anvil cirrus dynamics. \textit{Geophysical Research
  Letters}, \textbf{33}.

\bibitem[{Garrett et~al.(2005)}]{GarrettEtAl2005}
Garrett, T.~J., et~al., 2005: Evolution of a {F}lorida cirrus anvil. \textit{J.
  Atmos. Sci.}, \textbf{62}, 2352--2371, doi:10.1175/JAS3495.1.

\bibitem[{Haladay and Stephens(2009)}]{thinCirrus}
Haladay, T. and G.~Stephens, 2009: Characteristics of tropical thin cirrus
  clouds deduced from joint {C}loud{S}at and {CALIPSO} observations. \textit{J.
  Geophys. Res.}, \textbf{114}, D00A25, doi:10.1029/2008JD010\,675.

\bibitem[{Hartmann and Larson(2002)}]{HartmannAndLarson}
Hartmann, D.~L. and K.~Larson, 2002: An important constraint on tropical
  cloud-climate feedback. \textit{Geophysical Research Letters}, \textbf{29},
  1951--1955, doi:10.1029/2002GL015\,835.

\bibitem[{Herman(1980)}]{Herman1980}
Herman, G.~F., 1980: Thermal radiation in arctic stratus clouds. \textit{Quart.
  J. Roy. Meteor. Soc.}, \textbf{106}, 771--780.

\bibitem[{Jensen et~al.(2010)Jensen, Pfister, Bui, Lawson, and
  Baumgardner}]{Jensen2010TTL}
Jensen, E.~J., L.~Pfister, T.~P. Bui, P.~Lawson, and D.~Baumgardner, 2010: Ice
  nucleation and cloud microphysical properties in tropical tropopause layer
  cirrus. \textit{Atmos. Chem. Phys.}, \textbf{10}, 1369--1384,
  doi:10.5194/acp--10--1369--2010.

\bibitem[{Jensen et~al.(2011)Jensen, Pfister, and
  Toon}]{JensenTTLCirrusDynamics}
Jensen, E.~J., L.~Pfister, and O.~B. Toon, 2011: Impact of radiative heating,
  wind shear, temperature variability, and microphysical processes on the
  structure and evolution of thin cirrus in the tropical tropopause layer.
  \textit{J. Geophys. Res.}, \textbf{116}, D12\,209,
  doi:10.1029/2010JD015\,417.

\bibitem[{Jensen et~al.(1996)Jensen, Toon, Selkirk, Spinhirne, and
  Schoeberl}]{JensenEtAl96}
Jensen, E.~J., O.~B. Toon, H.~B. Selkirk, J.~D. Spinhirne, and M.~R. Schoeberl,
  1996: On the formation and persistence of subvisible cirrus clouds near the
  tropical tropopause. \textit{J. Geophysical Research}, \textbf{101},
  21,361--21,375.

\bibitem[{Jones et~al.(1986)Jones, Pyle, Harries, Zavody, III, and
  Gille}]{JonesEtAl}
Jones, R.~L., J.~A. Pyle, J.~E. Harries, A.~M. Zavody, J.~M.~R. III, and J.~C.
  Gille, 1986: The water vapour budget of the stratosphere studied using {LIMS}
  and {SAMS} satellite data. \textit{Quart. J. Roy. Meteor. Soc.},
  \textbf{112}, 1127--1143.

\bibitem[{Knollenberg et~al.(1993)Knollenberg, Kelly, and
  Wilson}]{Knollenberg93}
Knollenberg, R.~G., K.~Kelly, and J.~C. Wilson, 1993: Measurements of high
  number densitied of ice crystals in the tops of tropical cumulonumbus.
  \textit{J. Geophys. Res.}, \textbf{98}, 8639--8664.

\bibitem[{Lilly(1988)}]{Lilly88}
Lilly, D.~K., 1988: Cirrus outflow dynamics. \textit{J. Atmos. Sci.},
  \textbf{45}, 1594--1605.

\bibitem[{Liou et~al.(1988)Liou, Fu, and Ackerman}]{Liou88}
Liou, K.~N., Q.~Fu, and T.~P. Ackerman, 1988: A simple formulation of the
  delta-four-stream approximation for radiative transfer parameterizations.
  \textit{J. Atmos. Sci.}, \textbf{45}, 1940--1947.

\bibitem[{Mace et~al.(2006)Mace, Deng, Soden, and Zipser}]{cloudLifetime}
Mace, G.~G., M.~Deng, B.~Soden, and E.~Zipser, 2006: Association of tropical
  cirrus in the 10-15-km layer with deep convective sources: An observational
  study combining millimiter radar data and satellite-derived trajectories.
  \textit{J. Atmos. Sci}, \textbf{63}, 480--503, doi:10.1175/JAS3627.1.

\bibitem[{Raymond and Rotunno(1989)}]{Raymond&Rotunno89}
Raymond, D.~J. and R.~Rotunno, 1989: Response of a stably stratified flow to
  cooling. \textit{J. Atmos. Sci.}, \textbf{46}, 2830--2837.

\bibitem[{Scorer(1963)}]{CloudNomenclature}
Scorer, R.~S., 1963: Cloud nomenclature. \textit{Quart. J. Roy. Meteor. Soc.},
  \textbf{89}, 248--253.

\bibitem[{Spinhirne et~al.(1998)Spinhirne, Hart, and Duda}]{contrailDensity}
Spinhirne, J.~D., W.~D. Hart, and D.~P. Duda, 1998: Evolution of the morphology
  and microphysics of contrail cirrus from airborne remote sensing.
  \textit{Geophys. Res. Lett.}, \textbf{25}, 1153--1156,
  doi:10.1029/97GL03\,477.

\bibitem[{Toon et~al.(2010)}]{TC4}
Toon, O.~B., et~al., 2010: Planning, implementation, and first results of the
  tropical composition, cloud and climate coupling experiment (tc4). \textit{J.
  Geophys. Res.}, \textbf{115}, D00J04, doi:10.1029/2009JD013\,073.

\bibitem[{Voigt et~al.(2010)}]{CirrusSize}
Voigt, C., et~al., 2010: In-situ observations of young contrails - overview and
  selected results from the {CONCERT} campaign. \textit{Atmos. Chem. Phys.},
  \textbf{10}, 9039--9056, doi:10.5194/acpd--10--12\,713--2010.

\bibitem[{Zulauf(2001)}]{Zulauf}
Zulauf, M.~A., 2001: Modelling the effects of boundary layer circulations
  generated by cumulus convection and leads on large scale surface fluxes.
  \textit{Ph.D. Thesis, The Univesity of Utah}.

\end{thebibliography}

%
\end{document}